\documentclass[twocolumn]{aastex631}
\usepackage{epstopdf}
\usepackage{amssymb}
\usepackage{lineno}

\received{\today}
\revised{\today}
\accepted{\today}

\shorttitle{}
\shortauthors{Dai et al.}

\begin{document}

\title{Simultaneous Horizontal and Vertical Oscillation of a Quiescent Filament observed by CHASE and SDO}

\author[0000-0003-4787-5026]{Jun Dai}
\affiliation{Key Laboratory of Dark Matter and Space Astronomy, Purple Mountain Observatory, CAS, Nanjing, 210023, China}

\author[0000-0003-4078-2265]{Qingmin Zhang}
\affiliation{Key Laboratory of Dark Matter and Space Astronomy, Purple Mountain Observatory, CAS, Nanjing, 210023, China}
\affiliation{Yunnan Key Laboratory of the Solar physics and Space Science, Kunming, 650216, China}

\author[0000-0002-1190-0173]{Ye Qiu}
\affiliation{School of Astronomy and Space Science, Nanjing University, Nanjing 210023, China}
\affiliation{Key Laboratory for Modern Astronomy and Astrophysics (Nanjing University), Ministry of Education, Nanjing 210023, China}

\author[0000-0001-7693-4908]{Chuan Li}
\affiliation{School of Astronomy and Space Science, Nanjing University, Nanjing 210023, China}
\affiliation{Key Laboratory for Modern Astronomy and Astrophysics (Nanjing University), Ministry of Education, Nanjing 210023, China}

\author[0000-0002-4230-2520]{Zhentong Li}
\affiliation{Key Laboratory of Dark Matter and Space Astronomy, Purple Mountain Observatory, CAS, Nanjing, 210023, China}

\author[0000-0003-2694-2875]{Shuting Li}
\affiliation{Key Laboratory of Dark Matter and Space Astronomy, Purple Mountain Observatory, CAS, Nanjing, 210023, China}
\affiliation{School of Astronomy and Space Science, University of Science and Technology of China, Hefei, 230026, China}

\author[0000-0001-9647-2149]{Yingna Su}
\affiliation{Key Laboratory of Dark Matter and Space Astronomy, Purple Mountain Observatory, CAS, Nanjing, 210023, China}
\affiliation{School of Astronomy and Space Science, University of Science and Technology of China, Hefei, 230026, China}

\author[0000-0002-5898-2284]{Haisheng Ji}
\affiliation{Key Laboratory of Dark Matter and Space Astronomy, Purple Mountain Observatory, CAS, Nanjing, 210023, China}
\affiliation{School of Astronomy and Space Science, University of Science and Technology of China, Hefei, 230026, China}

\begin{abstract}

   In this paper, we present the imaging and spectroscopic observations of the simultaneous horizontal and vertical large-amplitude oscillation of a quiescent filament triggered by an EUV wave on 2022 October 02.
   Particularly, the filament oscillation involved winking phenomenon in H$\alpha$ images and horizontal motions in EUV images.
   Originally, a filament and its overlying loops across AR 13110 and 13113 erupted with a highly inclined direction, resulting in an X1.0 flare and a non-radial CME.
   The fast lateral expansion of loops excited an EUV wave and the corresponding Moreton wave propagating northward.
   Once the EUV wavefront arrived at the quiescent filament, the filament began to oscillate coherently along the horizontal direction
   and the ``winking filament" appeared concurrently in H$\alpha$ images.
   The horizontal oscillation involved an initial amplitude of $\sim$10.2 Mm and a velocity amplitude of $\sim$46.5 km s$^{-1}$,
   lasting for $\sim$3 cycles with a period of $\sim$18.2 minutes and a damping time of $\sim$31.1 minutes.
   The maximum Doppler velocities of the oscillating filament are 18 km s$^{-1}$ (redshift) and $-$24 km s$^{-1}$ (blueshift),
   which was derived from the spectroscopic data provided by CHASE/HIS.
   The three-dimensional velocity of the oscillation is determined to be $\sim$50 km s$^{-1}$ at an angle of $\sim50^\circ$ to the local photosphere plane.
   Based on the wave-filament interaction, the minimum energy of the EUV wave is estimated to be $2.7\times10^{20}$ J.
  Furthermore, this event provides evidence that Moreton waves should be excited by the highly inclined eruptions.

\end{abstract}


\correspondingauthor{Jun Dai and Qingmin Zhang}
\email{daijun@pmo.ac.cn, zhangqm@pmo.ac.cn}

\section{Introduction} \label{sec:intro}

The oscillatory phenomena in filaments are ubiquitous and complex dynamics, and have received more attention since the first systematic investigation performed by \cite{Hyder1966}.
Nowadays, based on the observed velocity amplitude \citep{Oliver2002,ball06}, the oscillatory motions in filaments are classified into large-amplitude oscillations (LAOs, $\geq$20 km s$^{-1}$) and small-amplitude oscillations (SAOs, $\leq$10 km s$^{-1}$).
Furthermore, the LAOs in filaments can be divided into two groups according to the direction, that is large-amplitude longitudinal oscillations \citep[LALOs,][]{jing03,jing06,vrs07,LZ12,zhang12,luna14,bi14,zhang17b,zhang20} where the filament material almost oscillate along its spine, and large-amplitude transverse oscillations \citep[LATOs, e.g.,][]{Isobe2007,pinter2008,chen08} where the filament considered as an long uniform cylinder oscillating perpendicular to its axis.
According to the recent reviews \citep{Oliver2009,Tripathi2009SSRv,arr18} and statistical study \citep{luna18}, both LATOs and LALOs are usually damping and lasting for several cycles, while the amplitudes and periods of LATOs are generally shorter than those of LALOs.
Particularly, transverse and longitudinal oscillations can occur simultaneously in a filament or prominence  \citep{pant2016,wang2016,zhang17a,luna21,dai2021}.

Generally, the LATOs are often triggered by Moreton waves \citep[e.g.,][]{Eto2002,Liu2013} or extreme-ultraviolet (EUV) waves \citep{liuw2012,daiy2012,Gosain2012,shen14a,shen2017,zhang2018}.
The restoring force of the LATOs is mainly supplied by the magnetic tension force of the corresponding filament, and the energy loss or dissipative process is theoretically considered as the reasonable damping mechanism \citep{Kleczek1969SoPh}.
Specifically, when the LATOs are observed on the limb, the line-of-sight (LOS) and plane-of-sky (POS) motions correspond to the horizontal and vertical oscillations,
while on the disk, the LOS and POS motions correspond to the nearly vertical and horizontal oscillations, respectively \citep{Tripathi2009SSRv}.
The vertical oscillations are usually associated with a phenomenon known as ``winking filament"  which are frequently observed on the disk in the earlier research \citep{Hyder1966,Ramsey1966}.
The winking filament are characterized with the alternative appearance and disappearance in H$\alpha$ wings images, due to the velocity component along the line of sight resulting in the Doppler shift from the H$\alpha$ center to the line wings \citep{Eto2002,Okamoto2004}.
The LOS velocity could be roughly derived from the simultaneous images in several wavelength channels close to the H$\alpha$ line-center
\citep[e.g., $\pm$0.4{~\AA} or $\pm$0.8{~\AA,}][]{Mor2003} and the value obtained in the previous studies ranged from 6 km s$^{-1}$ to 50 km s$^{-1}$ \citep{Isobe2006,Gilbert2008,Asai2012ApJ,shen14a}.

As the prime triggering mechanism for the LATOs, EUV waves are globally propagating wavelike phenomena characterized with brightening wavefronts followed by extending coronal dimming, which were first detected in 195{~\AA} \citep{Moses1997SoPh,Thompson1998GeoRL} by Extreme-ultraviolet Imaging Telescope \citep[EIT;][]{Delaboudini1995SoPh} on board the Solar and Heliospheric Observatory \citep[SOHO;][]{Domingo1995SoPh}. Now it is generally accepted that EUV waves are driven by the lateral expansion of coronal mass ejections (CMEs) flank rather than the pressure pulse of flares \citep{Biesecker2002ApJ,Cliver2005ApJ,Chen2006ApJ,Patsourakos2012SoPh}.
Theoretically, Moreton waves \citep{moreton1960,morram1960} are excited by the downward enhanced pressure when the coronal EUV waves front sweep over the underlying chromospheric plasma \citep{vrs2002,vrs2008}.
More recently, after investigating all the coronal-Moreton wave events over the past decade, \cite{zheng2023} uncovered that Moreton waves are more likely to be excited by the highly inclined eruptions.

Benefiting from the high-cadence data provided by the Atmospheric Imaging Assembly \citep[AIA;][]{lemen12} on board the Solar Dynamics Observatory \citep[SDO;][]{Pesnell2012}, an increasing number of EUV waves events containing both fast and slow components were reported \citep{chenwu2011ApJ,White2012AAS,daiy2012,Asai2012ApJ,chengxin2012ApJ,Kumar2013SoPh,shen2013ApJ,xue2013,zongdai2015,liurui2019ApJ}.
The hybrid wave theory \citep{chenpf2002} where the fast component was considered as the fast-mode MHD wave or shock wave \citep{Thompson1998GeoRL,wang2000ApJ,Wu2001JGR} and the slow component was supposed as the apparent wave induced by the stretching of magnetic field lines, is thus well supported by the thermodynamic MHD simulation \citep{Downs2012ApJ} and observations \citep{sun2022,Chenpf2023}.
Typically, the fast-component waves present outer sharp wavefronts and propagate with speed from $\sim$500 km s$^{-1}$ to more than 1500 km s$^{-1}$, while the slow-component waves present inner diffuse wavefronts with speeds usually about one-third of the fast-component waves, sometimes even propagating slower than the coronal sound speed \citep{Tripathi2007A&A,Thompson2009ApJS,Zhukov2009SoPh}.
However, the physical nature of EUV waves is still under controversy and more details is available in recent reviews \citep{Liu2014SoPh,Warmuth2015LRSP,chen2016GMS,Shen2020ChSBu,Chenpf2023}.

On 2022 October 02, an X1.0 flare occurred in the active region (AR) 13110 and 13113, which was accompanied by an EUV wave (Moreton wave) and a non-radial CME.
The heating mechanisms related to the flare have been reported by \cite{song2023}.
In this paper, based on the imaging and spectroscopic observations, we focus on the simultaneous POS horizontal oscillation and winking phenomenon of a quiescent filament triggered by the EUV wave and Moreton wave, which was rarely reported in  pervious studies.
The observation and data analysis are described in Section~\ref{sec:obs}.
The results are presented in Section~\ref{sec:res}.
A comparison with previous findings and a brief conclusion are given in Section~\ref{sec:sum}.

\begin{figure*}
\centering\includegraphics[width=0.85\textwidth]{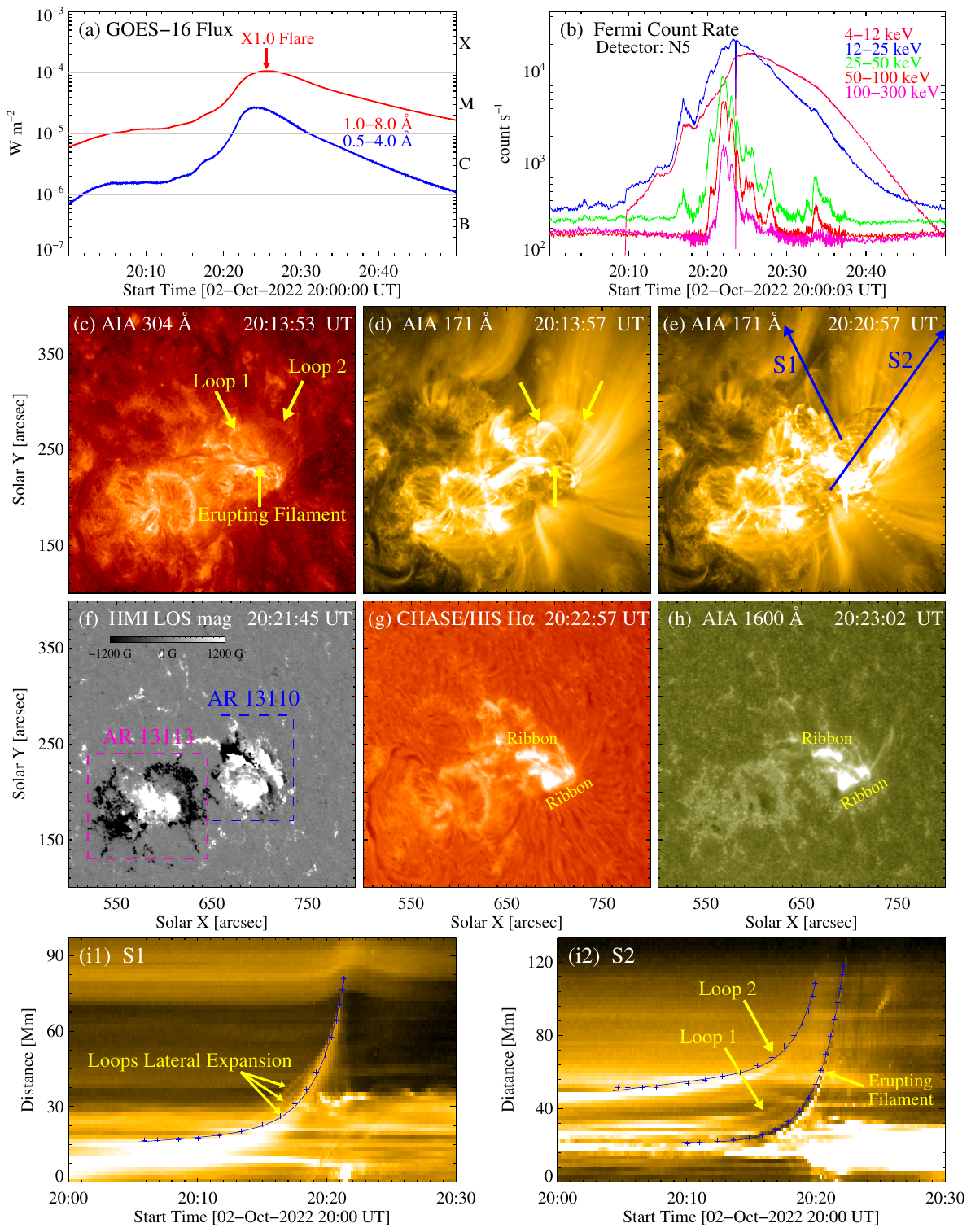}
\caption{\small{The overview of the eruptions and the X1.0 flare in AR 13110 and 13113.
Panel (a): SXR light curves of the X1.0 flare in 1-8{~\AA} (red line) and 0.5-4{~\AA} (blue line);
Panel (b): HXR fluxes of the X1.0 flare at various energy bands (4-300 keV);
Panels (c-e) give three EUV images in AIA 304{~\AA} and 171{~\AA}.
Three yellow arrows point to erupting filament and loops, respectively;
Panel (f) give an HMI LOS magnetogram with the magnetic field strengths in the range of -1200 and 1200 G;
Panels (g-h) provide an H$\alpha$ image observed by CHASE and a UV image in AIA 1600{~\AA}.
Panels (i1-i2): Time-distance diagrams of the slices S1 and S2 in AIA 171{~\AA}.
The blue diamonds outline the heights of the erupting filament and loops, and the blue solid lines represent the fitted curve using Equation (1).}}
\label{fig1}
\end{figure*}

\section{Observation and data analysis} \label{sec:obs}

\subsection{Data and Instruments}

On 2022 October 02, AR 13110 (N17W52) and 13113 (N16W43) were adjacent and located near the western limb.
The X1.0 flare, the EUV wave (Moreton wave) and the transverse oscillation of the quiescent filament in this study were simultaneously recorded by H$\alpha$ Imaging Spectrograph (HIS) on board the Chinese H$\alpha$ Solar Explorer \citep[CHASE;][]{Lichuan2022SCPMA} and Atmospheric Imaging Assembly \citep[AIA;][]{lemen12} on board the Solar Dynamics Observatory \citep[SDO;][]{Pesnell2012}. Furthermore, it should be noted that the observation of CHASE for this event was only from 20:15 UT to 20:41 UT.

 As the first space solar mission of China, CHASE was lunched into a Sun-synchronous orbit on 2021 October 14.  HIS takes full-disk spectroscopic observations at H$\alpha$ (6559.7-6565.9{~\AA}) and Fe \small I \normalsize(6567.8-6570.6{~\AA}) wavebands in raster scanning mode (RSM), with a spectral resolution of 0.024{~\AA} pixel$^{-1}$, a spatial resolution of 0$\farcs$52 pixel$^{-1}$  and a temporal resolution of 60 s.
The detailed calibration procedures from the raw data to high-level products involved the dark field correction, flat field and slit image curvature correction, wavelength and intensity calibration, and coordinate transformation \citep{Qiuye2022SCPMA}. We employed both the images at H$\alpha$ line center (6562.82{~\AA}), H$\alpha$ wings (i.e., $\pm$0.4{~\AA}, $\pm$0.8{~\AA}), and the derived Dopplergrams to study the motions of the oscillating filament and the Moreton wave.

AIA provides full-disk images in two ultraviolet (UV; 1600 and 1700{~\AA}) wavelengths with a cadence of 24 s and in seven EUV (94, 131, 171, 193, 211, 304, and 335{~\AA}) wavelengths with a cadence of 12 s,
we mainly employed the EUV images to study the dynamics of the EUV wave.
The photospheric LOS magnetograms of AR 13110 and 13113 were observed by the Helioseismic and Magnetic Imager \citep[HMI;][]{schou12} on board SDO with a cadence of 45 s.
The level\_1 data from AIA and HMI with a spatial resolution of 1$\farcs$2 were calibrated using the standard Solar SoftWare (SSW) programs \texttt{aia\_prep.pro} and \texttt{hmi\_prep.pro}.

In addition, soft X-ray (SXR) light curves of the flare in 0.5$-$4{~\AA} and 1$-$8{~\AA} were recorded by the GOES-16 spacecraft with a cadence of 2 s, and hard X-ray (HXR) fluxes at various energy bands (4-300 keV) were detected by the Gamma-ray Burst Monitor \citep[GBM;][]{Meegan2009ApJ} on board the Fermi spacecraft.
The erupting filament across the two active regions and the related non-radial CME were also observed by the Extreme-UltraViolet Imager \citep[EUVI;][]{Wuelser2004SPIE} and the COR2 white-light coronagraph on board the Ahead Solar Terrestrial Relation Observatory \citep[STEREO-A;][]{Kaiser2008SSRv}.
The separation angle between STEREO-A and Earth was $\sim$17$^\circ$.

\subsection{Models and Methods}

The cloud model \citep{Beckers1964,Mein1988,chae2014,hong2014} is widely used to indirectly derive the 3D velocity of erupting or oscillating filaments based on H$\alpha$ observations \citep{Mor2003,Mor2010,shen14a}.
To obtain the Doppler velocity, we first define the line center of the averaged spectral profile in a quiescent region [600$\arcsec$, 615$\arcsec$]$\times$[525$\arcsec$, 535$\arcsec$] as zero shift to avoid the influence of solar rotation and instrument movement. Then, we recognize the filament region by larger line depth and lower intensity. At the filament region, we adopt the single cloud model  to acquire the line-of-sight velocity of filament. For the other region without filament, the line-of-sight velocity is calculated by the moment analysis \citep{Li2019,Yu2020}.

The revised cone model is developed to track the 3D evolution of filament eruptions or CMEs and investigate the properties of the non-radial filament eruptions or CMEs, which require observations from at least two viewpoints \citep{zhang2021A&A,zhang2022A&A}.
The revised cone model is based on the assumption that the shape of the eruptions or CMEs is the ice-cream cone, where the cone apex is located at the source regions and the cone base is spheric.
The revised cone model is characterized by the matrix transforms between the heliocentric coordinate system (HCS), local
coordinate system (LCS) and cone coordinate system (CCS) to determine the geometric parameters, such as the inclination angle $\theta_1$, $\phi_1$ and angular width $\omega$.
To construct the revised cone model, we select the images simultaneously observed by the STA/EUVI 195{~\AA} and SDO/AIA 193{~\AA} at 20:22 UT.

\section{Result} \label{sec:res}

\subsection{X1.0 flare and Eruptions}

In Figure~\ref{fig1}, the top panels show the SXR (1-8{~\AA} and 0.5-4{~\AA}) and HXR (4-300 keV) fluxes of the X1.0 flare in AR 13110 and 13113, which peaked at $\sim$20:25 UT and $\sim$20:22 UT, respectively.
It is worthy noting that the X1.0 flare was associated with eruptions of a filament and its overlying loops crossing AR 13110 and 13113,
which are marked by three yellow arrows in Figure~\ref{fig1}(c-d).
From $\sim$20:05 UT to $\sim$20:22 UT, the overlying loops and the filament erupted successively,
resulting in two roughly parallel flare ribbons denoted by yellow marks in Figure~\ref{fig1}(g-h).

To investigate the kinematics of the eruptions,
we select two long straight slices S1 and S2 with lengths of 97 Mm and 133 Mm in Figure~\ref{fig1}(e) to construct the time-space diagrams in AIA 171{~\AA} of Figure~\ref{fig1}(i1-i2), since the eruptions were essentially in the same direction along S2 while the lateral expansion of loops moved along S1.
As shown in Figure~\ref{fig1}(i1-i2), the signals of the loops lateral expansion and eruptions pointed by yellow arrows can be clearly recognized.
The height of Loop 1 lateral expansion along S1, as well as the heights of Loop 2 and filament eruptions along S2 can be obtained by outlining the edge in the diagrams.
The height evolution in the plane-of-sky is characterized by a slow rise with a constant speed followed by an initial impulsive acceleration indicated by an apparent exponential increase. Hence, we fit $h(t)$ using the function composed of a linear term and an exponential term as proposed by \cite{cheng13}:
\begin{equation} \label{eqn-1}
  h(t) = c_{0}e^{(t-t_{0})/{\tau}} + c_{1}(t-t_{0}) + c_{2},
\end{equation}
where $t_0$, $\tau$, $c_0$, $c_1$, and $c_2$ are free parameters.
The onset of fast-rise phase is defined by setting the parameters as the point when the exponential velocity is equal to the linear velocity:
\begin{equation} \label{eqn-2}
  t_{onset} = \tau\ln(c_1 \tau/c_0)+t_0.
\end{equation}
The fitted curve is plotted with the blue solid line in Figure~\ref{fig1}(i1-i2).
Precisely, loops started to rise slowly from $\sim$20:05 UT while their fast rise and lateral expansion started at $\sim$20:18 UT.
The filament started to rise slowly from $\sim$20:10 UT and began to rise fast at $\sim$20:20 UT.

\begin{figure*}[htb]
\centering\includegraphics[width=0.99\textwidth]{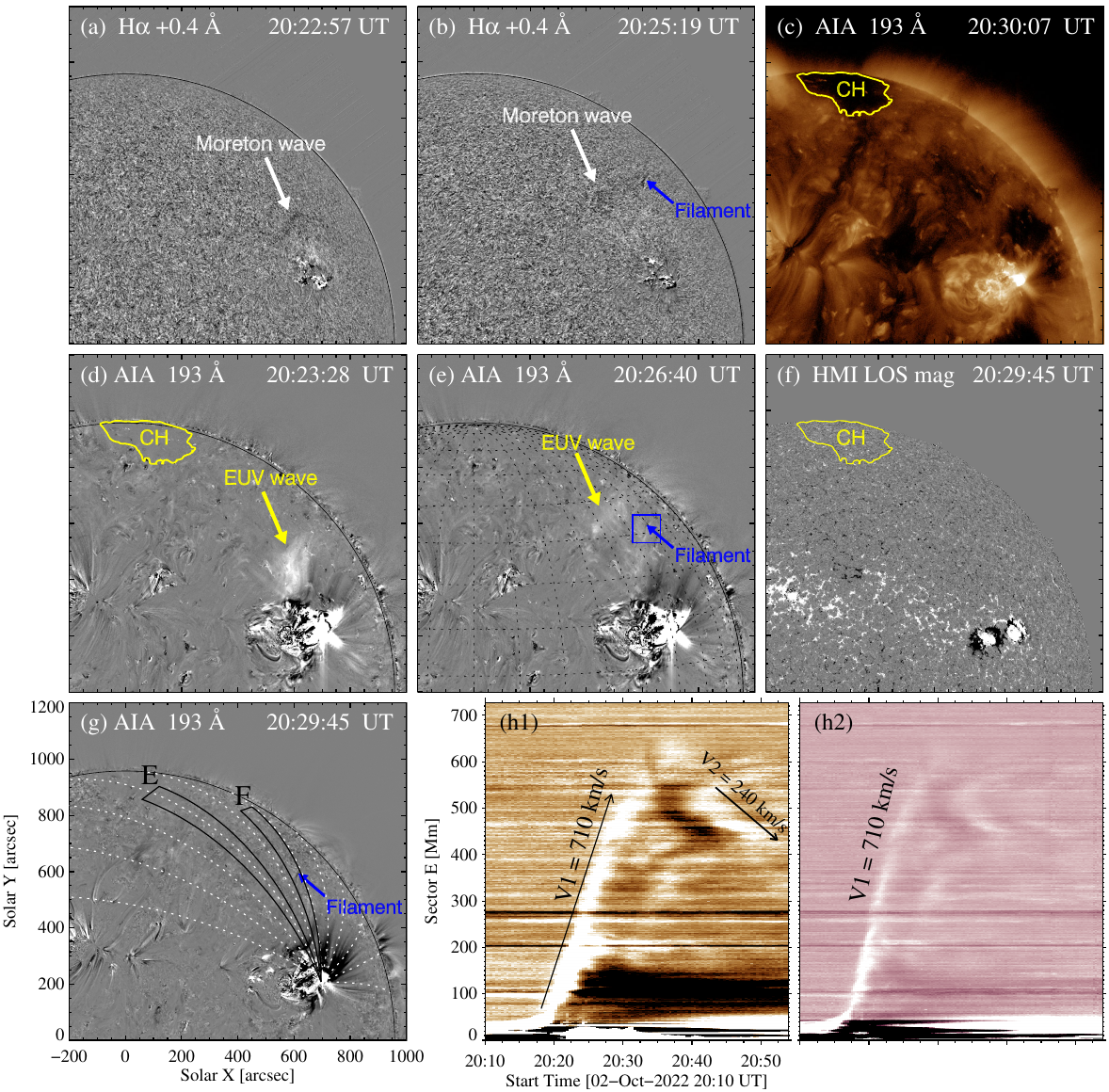}
\caption{\small{ Kinematics of EUV wave and Moreton wave.
Panels (a-b): Running differences of H$\alpha$ $+$0.4{~\AA} images.
Panel (c): An AIA 193{~\AA} image.
Panels (d,e,g) give three AIA 193{~\AA} base-difference images.
Panel (f): An HMI LOS magnetogram.
   Panel (h1-h2) provide time-distance diagrams along the sector E in panel (g) using the 193{~\AA} and 211{~\AA} base-difference images.
   The Moreton wave, the EUV wave and the oscillating filament are separately marked by white, yellow and blue arrows.
   The yellow contours in panels (c,d,f) mark the coronal hole.
   The white dotted lines in panel (g) partially display the longitude of the new solar coordinate system centered on the X1.0 flare, denoting the same sector as the EUV wave.
   The propagation of EUV wave through the oscillating filament is marked by sector F.
   The speeds of the propagating EUV wave are denoted by numbers in panels (h1-h2).
   The blue rectangle in panel (e) shows the FOV of Figure~\ref{fig4}(a)-(c) and Figure~\ref{fig6}(a).}
   An animation of the unannotated SDO observations is available that covers $\sim$45 minutes starting at 20:10 UT and ending the same day at 20:55 UT, with a time cadence of 24 seconds.
   (An animation of this figure is available.)}
\label{fig2}
\end{figure*}

\begin{figure*}[htb]
\centering\includegraphics[width=0.99\textwidth]{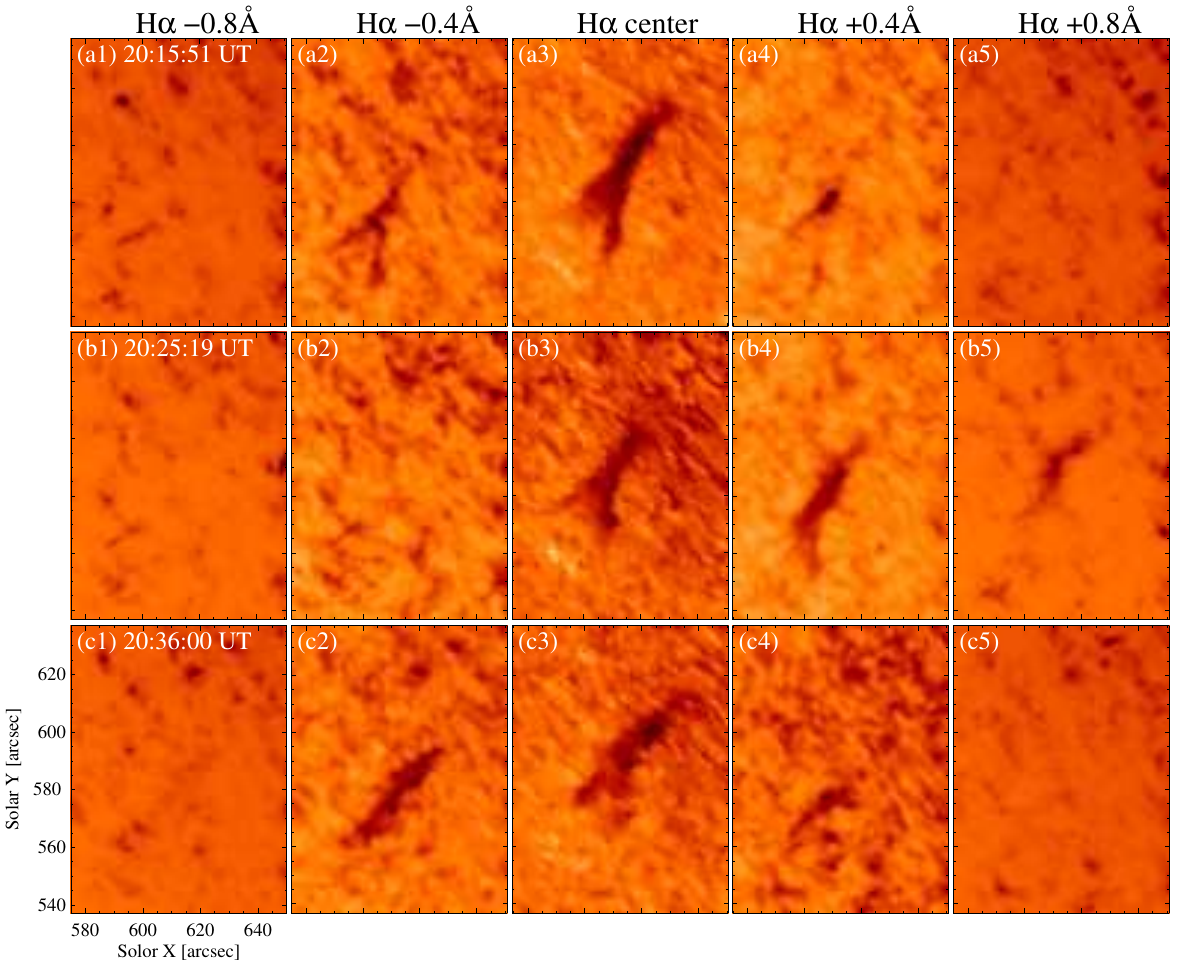}
\caption{\small{Sequential CHASE/HIS H$\alpha$ images showing the oscillating filament.
The middle column presents the images of H$\alpha$ line-center (6562.82{~\AA}), while its left and right columns are images of two H$\alpha$  blue wings ($-$0.8{~\AA}, $-$0.4{~\AA})
    and two H$\alpha$ red wings ($+$0.4{~\AA}, $+$0.8{~\AA}), respectively.
    Images in the same row are recorded at the same time.
  An animation of the unannotated SDO observations is available that covers $\sim$26 minutes starting at 20:15 UT and ending the same day at 20:41 UT, with a time cadence of 1 minute.
  (An animation of this figure is available.)}}
\label{fig3}
\end{figure*}

\subsection{Triggering and Kinematics of EUV wave and Moreton wave}

Figure~\ref{fig2} presents the propagation of the Moreton wave in running difference images in H$\alpha$ $+$0.4{~\AA} and the EUV wave in base difference images in 193{~\AA}.
Combining the online animation of Figure~\ref{fig2}(d-e) and the fitted curve in Figure~\ref{fig1}(i1), it can be seen that the onset of the fast lateral expansion of loops was cotemporal with the first trace of the EUV wave at $\sim$20:18 UT,
implying that the EUV wave was excited by the fast loops lateral expansion \citep{Patsourakos2009ApJ,Patsourakos2010AA}.
Subsequently, the EUV wave propagated northward in a fan shape centered on the AR 13110 and its bright segment is pointed by yellow arrows in Figure~\ref{fig2}(d-e), while the simultaneous Moreton-wave dark front is pointed by white arrows in Figure~\ref{fig2}(a-b).
What is noteworthy is that the bright segment was reflected at the edge of the polar coronal hole \citep{Gopalswamy2009}, which is marked by yellow contours in Figure~\ref{fig2}.

In order to analyze the kinematics of the EUV wave, we select a sector (E) along the propagating direction of the bright segment with a width of 10$^\circ$ and a length of 726 Mm in Figure~\ref{fig2}(g), and the corresponding time-space diagrams of 193{~\AA} and 211{~\AA} base-difference images are plotted in Figure~\ref{fig2}(h1-h2).
It can be seen that the bright segment of the EUV wave propagated at a speed of $\sim$710 km s$^{-1}$ from $\sim$20:18 UT.
The perpendicular reflection at the edge of the polar coronal hole occurred at $\sim$20:35 UT, then the bight segment traveled for $\sim$ 200 Mm at a speed of $\sim$240 km s$^{-1}$, after which the wavefront could not be distinguished in the EUV images.
The deceleration of the EUV wave after reflection here is consistent with the values in previous reports \citep{Olmedo2012,liT2012,shen2012}.

\begin{figure*}[htb]
\centering\includegraphics[width=0.9\textwidth]{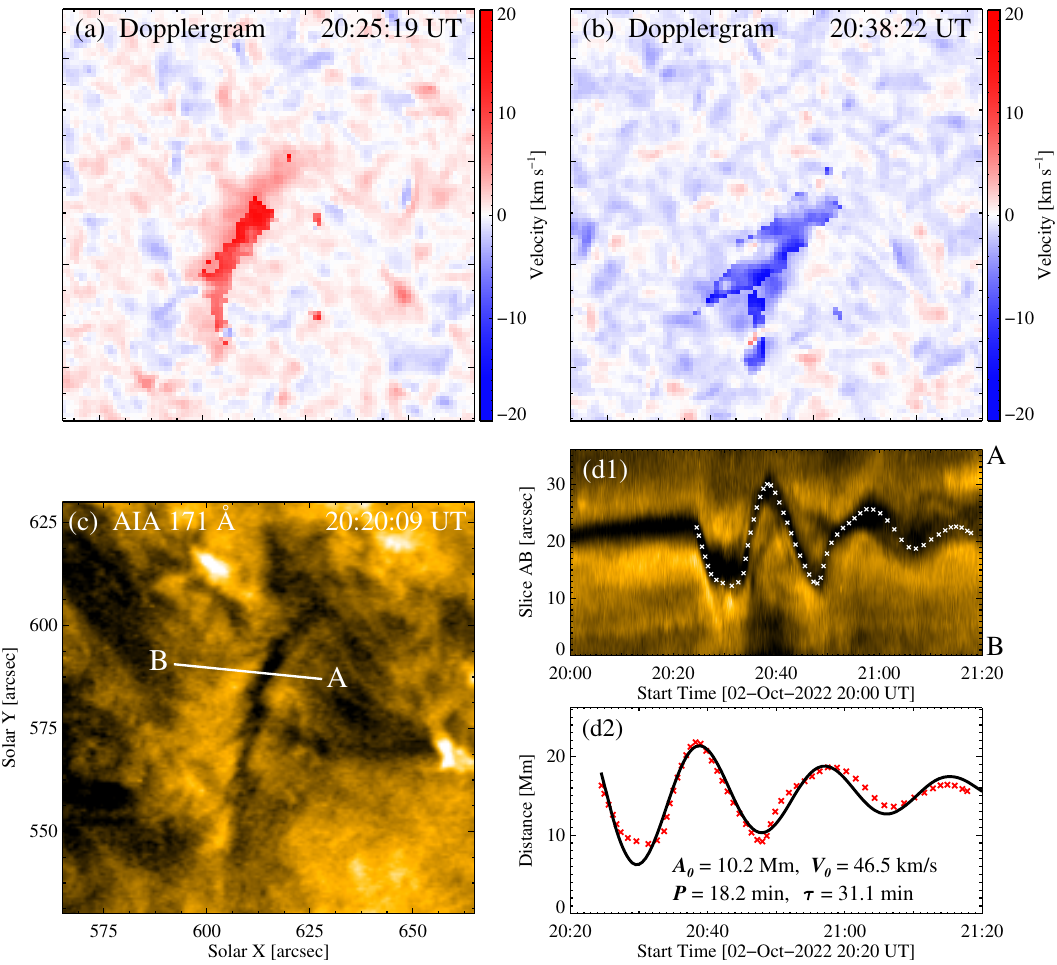}
\caption{\small{
Panels (a-b): Two Dopplergrams showing the velocities of the filament material along the LOS direction during the oscillation.
Panel (c): An EUV image in AIA 171{~\AA} of the filament before the oscillation.
Panel (d1): Time-distance diagrams of the slice AB in AIA 171{~\AA}.
The white crosses outline the filament oscillation between 20:20 UT and 21:20 UT.
Panel (d2): Extracted positions of the oscillating filament along the slice AB.
The fitted curve is overlaid with black solid line.
Corresponding parameters are labeled.
}}
\label{fig4}
\end{figure*}

\subsection{Simultaneous Horizontal and Vertical Oscillation}

A quiescent filament (N40W60) pointed by blue arrows in Figure~\ref{fig2} was located north of the two active regions at a distance of $\sim$260 Mm.
At $\sim$20:24 UT, the bright segment of the EUV wave swept over the quiescent filament, after which the winking phenomenon and the obvious horizontal oscillation of the filament was respectively detected in H$\alpha$ and EUV images, indicating that the oscillation contained both POS and LOS velocity components at the same time.

Figure~\ref{fig3} presents the time sequence of H$\alpha$ line-center (middle column), blue wing (left columns) and red wing (left columns) images to illustrate the upward and downward motions of the filament along the LOS.
The first row shows the filament features before oscillation, from which we can see in the H$\alpha$ line-center image that the filament consists of a spine and a barb on its left side, while the barb and the southern leg of the filament can be observed in $\pm$0.4{~\AA} images, probably due to the counter-streaming of plasma \citep{zirker1998}, along the LOS direction therein. At $\sim$20:25 UT, as shown in the second row in Figure~\ref{fig3}, the filament appeared a decrease in intensity with increasing wavelength in the red wing images, while disappearing completely in the blue wing images, indicating that the filament was first undergoing downward motion after being squeezed by the EUV wave.
After about 10 minutes, the filament underwent upward motion and only appeared in the blue wing images (see the third row and the online animation of Figure~\ref{fig3}).
The subsequent winking phenomenon of the filament could not be presented here due to the limitation of the observation time of CHASE, which only covered the first cycle of the oscillation.
Nevertheless, we obtain the Doppler velocities of the oscillating filament from the spectroscopic data of HIS and the maximum velocities are shown in Figure~\ref{fig4}(a-b), which were 18 km s$^{-1}$ (redshift) and $-$24 km s$^{-1}$ (blue shift), respectively.

In order to investigate the horizontal motions of the filament, we select a straight slice (AB) perpendicular to the axis with a length of 36$\arcsec$ in Figure~\ref{fig4}(c), and the corresponding time-space diagrams in 171{~\AA} are plotted in Figure~\ref{fig4}(d1).
It can be seen that the filament did not exhibit obvious oscillations from 20:00 UT to 20:23 UT.
At $\sim$20:24 UT, when the EUV waves arrived, the filament began to deviate from the equilibrium position and moved coherently eastward. Then, the filament moved backward and oscillated with decaying amplitude for about three cycles.
These phenomena indicate that the filament oscillation was triggered by the EUV wave, rather than the tornado phenomenon of the quiescent filament \citep{su2012,Panasenco2014}.
To accurately determine the parameters of the filament oscillation,
we fit the white curve in Figure~\ref{fig4}(d1) using \texttt{mpfit.pro} and the following exponentially decaying sine function:
\begin{equation} \label{eqn-3}
  A(t) = A_{0}\sin \bigg(\frac{2\pi t}{P} + \psi \bigg)e^{-\frac{t}{\tau}} + A_{1}t + A_{2},
\end{equation}
where $A_0$ is the initial amplitude, $P$ is the period, $\tau$ is the damping time, $\psi$ is the initial phase,
and $A_{1}t+A_{2}$ represents a linear term of the equilibrium position of the filament.
In Figure~\ref{fig4}(d2), the red crosses represent the extracted positions of the oscillating filament along the slice AB in 171{~\AA}, and the fitting result using Equation~\ref{eqn-1} was overlaid with black line.
As seen from the fitted parameters listed in Figure~\ref{fig4}(d2), the filament oscillation
has an initial amplitude of $\sim$10.2 Mm and a velocity amplitudes of $\sim$46.5 km s$^{-1}$. The period and the damping time are $\sim$18.2 minutes and $\sim$31.1 minutes, and the corresponding quality factor ($\tau/P$) is $\sim$1.7. Therefore, the quiescent filament underwent a large amplitude transverse oscillation (LATO).

\begin{figure}
\centering\includegraphics[width=0.475\textwidth]{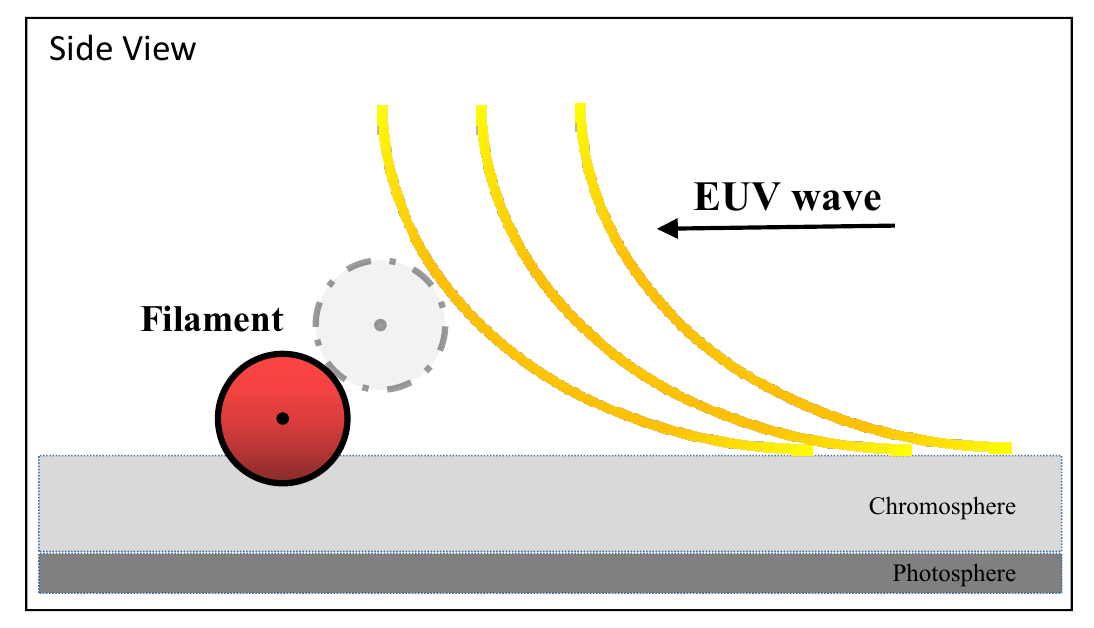}
\caption{\small{Schematic diagrams on side view showing the scenario of the wave-filament interaction.
The yellow curves represent the EUV wave and the black arrow indicates the propagation of the EUV wave.
The dashed gray and solid red circles indicate the position of filament before the oscillation and the position of the maximum displacement after the impact of the EUV wave, respectively. The black (gray) dots indicate the axis of the filament spine. The light and dark gray layer on the bottom represent chromosphere and photosphere, respectively.
} }
\label{fig5}
\end{figure}

Combining the initial velocity amplitudes (POS) and the Doppler velocity (LOS) around 20:25 UT, as well as the location of the quiescent filament (N40W60),
schematic diagrams showing the side view of the wave-filament interaction are provided in Figure~\ref{fig5},
the initial maximum displacement and 3D velocity of the oscillation is determined to be $\sim$12 Mm and $\sim$50 km s$^{-1}$, the direction of which was at an angle of $\sim50^\circ$ to the local photosphere plane.

According to the prominence seismology applied to the winking filaments \citep{pinter2008,shen14b}, the radial magnetic field strength of the quiescent filament can be expressed as the following equation \citep{Hyder1966} based on the measured parameters of the LATO:
\begin{equation} \label{eqn-4}
B_r^2 = \pi \rho \ r_0^2[\ 4\pi ^2(\frac{1}{P})^2+(\frac{1}{\tau})^2],
\end{equation}
where $B_r$ is the radial component of the filament magnetic field, $\rho=N_\mathrm{H}m_\mathrm{H}$ is the mass density in the filament ($N_\mathrm{H}$ is the mean neutral hydrogen number density and $m_\mathrm{H}$ is the mass of the hydrogen atom), $r_0$ is the scale height of the filament, $P$ is the period, and $\tau$ is the damping time.
If we set the value $r_0$ = $3\times10^{9}$ cm \citep{Hyder1966},
using the observed value of $P$ = 1092 s, $\tau$ = 1866 s,
and taking the value of $m_\mathrm{H} = 1.67\times10^{-24}$ g, $N_\mathrm{H} =3 \times10^{11}$ cm$^{-3}$ in the quiescent filaments \citep{Heinzel2008ApJ}, the radial component of the filament magnetic field is estimated to be about 6.5 Gauss, which is consistent with the values in previous studies \citep[e.g.,][]{shen14b}.

\begin{figure}
\centering\includegraphics[width=0.475\textwidth]{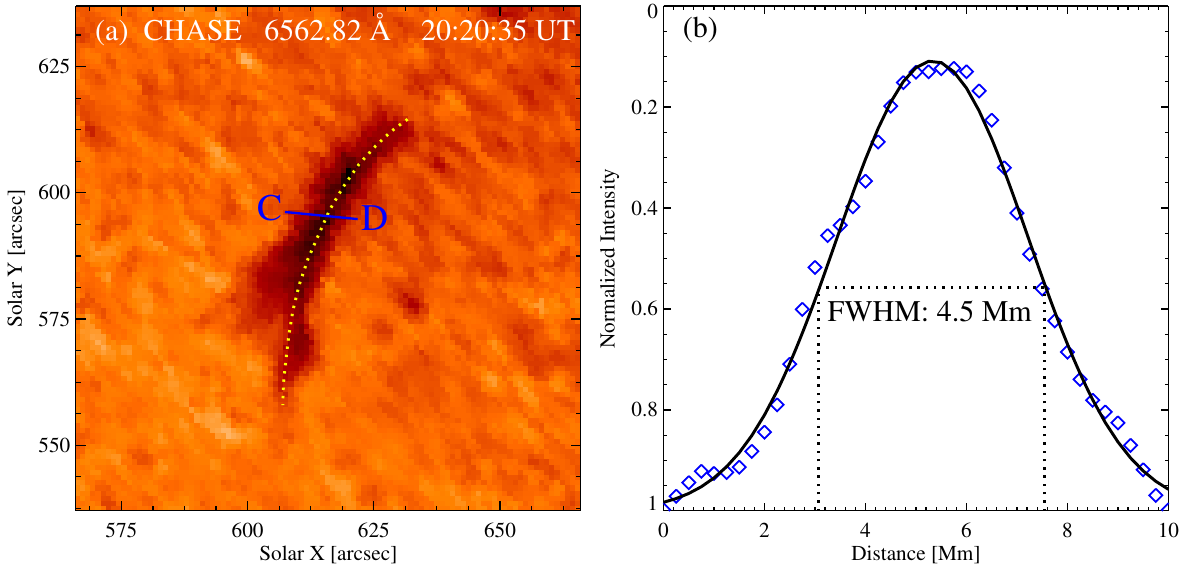}
\caption{\small{Panel (a): An H$\alpha$ image in HIS 6562.82{~\AA} of the filament before the oscillation. The spine of the filament is marked by a dotted yellow line.
Panel (b): Normalized intensity distribution along the line CD across the oscillating filament, which is fitted with a single-Gauss function. The full width at half maximum (FWHM) representing the width or LOS depth of the filament is labeled.
}}
\label{fig6}
\end{figure}

\subsection{Energetics of Filament Oscillation and EUV wave}

Since the transverse oscillation of the quiescent filament was triggered by the impact of EUV wave,
the kinetic energy transferred from the EUV wave to the filament can be estimated based on their interaction.
The mass of a filament can be estimated roughly using the following expression \citep{Labrosse2010SSRv}:
\begin{equation} \label{eqn-5}
 M = N_\mathrm{H}m_\mathrm{H}V,
\end{equation}
where $V$ represents the volume of the filament plasma.
Typical range of $N_\mathrm{H}$ in the quiescent filaments is
$3\times10^{10}-10^{11}$ cm$^{-3}$ at 6000$-$8000 K \citep{Heinzel2008ApJ,Labrosse2010SSRv}.
In the current study, as showing the Figure~\ref{fig6}, the apparent width and length of the oscillating filament are determined to be 4.5 Mm and 43.2 Mm, respectively.
Assuming the geometry occupied by the filament plasma is a uniform cylinder,
and considering that the filament is composed of thin threads with the filling factor of 0.1 \citep{lin2005,Martin2008,lin2009},
the total mass of the oscillating filament is estimated to be $3.5\times10^{12}-1.2\times10^{13}$ g.
Taking the 3D velocity ($\sim$50 km s$^{-1}$) of the oscillation and ensuring that the oscillation can be triggered in case of the maximum mass of the filament \citep{Gilbert2008}, 
the minimum energy transferred from the EUV wave to the filament is $1.5\times10^{19}$ J,
which is actually the initial kinetic energy of the oscillating filament.
As shown in the Figure~\ref{fig2}(g), the propagation of the EUV wave is
 denoted by the partial longitude of the new solar coordinate system with an angle of $\sim$165$^\circ$, 
 while the wave-filament interaction is marked by the sector F with an angle of $\sim$10$^\circ$.
Assuming that the sectoral propagation of the EUV wave is isotropic,
the area ratio between the filament and the EUV wave dome can be approximated as the angle ratio between the sector F and the propagating sector of the EUV wave.
Thus, the minimum energy of the EUV wave is estimated to be $2.7\times10^{20}$ J.
Furthermore, the minimum energy on the interaction between the EUV wave and coronal hole is estimated to be $1.8\times10^{20}$ J, 
since the velocity of the EUV wave was decelerated to one-third after the reflection.

\begin{figure*}[htb]
\centering\includegraphics[width=0.99\textwidth]{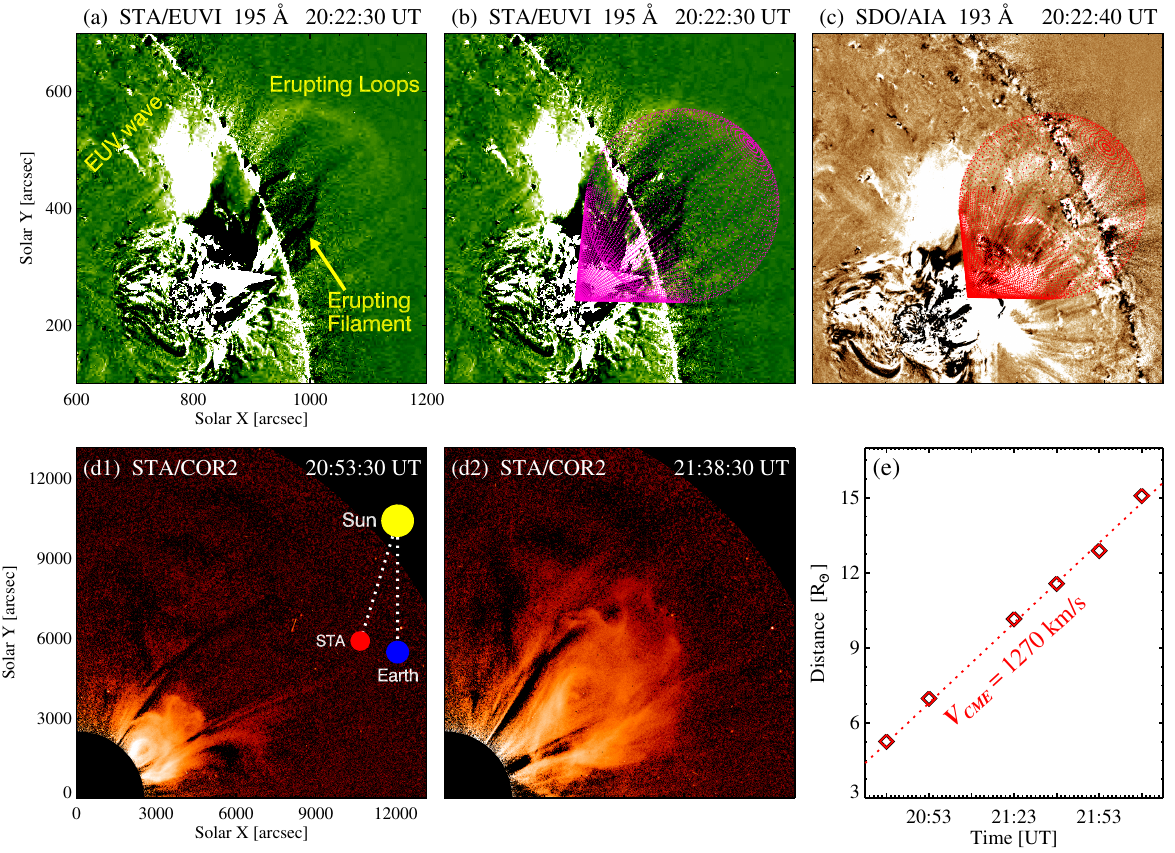}
\caption{\small{Panel (a): An STA/EUVI 195{~\AA} image of the erupting filament and loops.
Panels (b-c): The projections of the reconstructed cone are separately depicted by magenta and red dots, 
and overlaid on the simultaneous STA/EUVI 195{~\AA} and AIA 193{~\AA} base-difference images at 20:22 UT.
Panels (d1-d2): Two snapshots of the white-light CME observed by STA/CRO2. 
The relative positions of the Sun, Earth and STEREO-A at 20:53 UT on 2022 October 02 are plotted in panel (d1).
Panel (e): Height-time plot of the CME leading edges with red square symbols.
The results of a linear fit is overlaid with a dashed line and the linear speed is labeled.} }
\label{fig7}
\end{figure*}

\subsection{Non-radial loops eruption and CME}

As shown in Figure~\ref{fig7}, the eruptions across two active regions and the associated CME were obviously deviated northward from the radial direction.
To determine the geometrical parameters of the non-radial eruption, 
we select the images simultaneously observed by the STA/EUVI 195{~\AA} and SDO/AIA 193{~\AA} at 20:22 UT to construct the revised cone model, 
and the projections of the ice-cream cone are overlaid with magenta and red dots in SDO/AIA and STA/EUVI field of view, which are displayed in the top panels of Figure~\ref{fig7}.
It can be seen that the ice-cream cone fits well with the erupting loops.
The inclination angle $\theta_1$, $\phi_1$ and angular width $\omega$ obtained from the cone model are 24$^\circ$ (deviated northward), 5$^\circ$ (deviated eastward) and 80$^\circ$, respectively. 
Thus, the loops eruption was barely deviated from the plane of sky \citep[cf. the topology of the revised cone model and the transformations of coordinate systems from][]{zhang2021A&A}.
Hence, the linear velocity (1270 km s$^{-1}$) of the CME estimated from the STA/COR2 is approximately equal to its true speed.

\section{Summary and Discussion}\label{sec:sum}

In this paper, we report the simultaneous horizontal and vertical oscillation of a quiescent filament recorded by CHASE and SDO on 2022 October 02.
The filament oscillation was triggered by an EUV wave accompanied with Moreton wave, which was excited by a non-radial loops eruption in AR 13110 and 13113.
Particularly, the filament oscillation involved winking phenomenon in H$\alpha$ images and horizontal motions in EUV images.
Specifically, the quiescent filament underwent the POS horizontal oscillation with an initial amplitude of $\sim$10.2 Mm and a velocity amplitude of $\sim$46.5 km s$^{-1}$, 
lasting for $\sim$3 cycles with a period of $\sim$18.2 minutes and a damping time of $\sim$31.1 minutes.
The Doppler velocities computed from the CHASE/HIS spectroscopic data are 18 km s$^{-1}$ (redshift) and $-$24 km s$^{-1}$ (blueshift), respectively.
Combining the Doppler velocities and the location of the quiescent filament, we obtain the value and direction of the 3D velocities, 
which was $\sim$50 km s$^{-1}$ at an angle of $\sim50^\circ$ to the local photosphere plane.
According to the prominence seismology, the radial magnetic field strength of the quiescent filament is determined to be $\sim$6.5 Gauss.
Furthermore, the minimum energy of the EUV wave is estimated to be $2.7\times10^{20}$ J based on the wave-filament interaction.

The transverse (horizontal) and vertical oscillations of filaments have been widely reported and studied since 1930s \citep{Hyder1966,Tripathi2009SSRv,arr18}.
When the oscillating filament are located far from the center of the solar disk, 
there are projection effects for both horizontal and vertical oscillations \citep{Kleczek1969SoPh}, which can be removed by combining POS and LOS velocities.
However, the simultaneous horizontal oscillation and winking phenomenon in a filament was rarely reported in previous studies.
To our knowledge, only \cite{Isobe2006} and \cite{Gilbert2008} reported the similar filament oscillations which involved both POS horizontal and LOS vertical motions, 
while the accurate 3D velocities were not determined due to the limitations of data.
Although there was no winking filament observed \citep{Isobe2006}, the Doppler velocities derived from two H$\alpha$ wings ($\pm$0.8{~\AA}) images was up to 50 km s$^{-1}$, 
which was obviously deviated from the observation.
Likewise, based on a few available wavelengths (H$\alpha$ and He \small I \normalsize 10830{~\AA}), the LOS velocity component was estimated to be 41 km s$^{-1}$ \citep{Gilbert2008}, 
which was also controversial since the Doppler velocity obtained by \cite{Jack2013} was only a few km s$^{-1}$ in their reanalysis for the same event.
Using the images in H$\alpha$ line-center and six H$\alpha$ wings ($\pm$0.5{~\AA}, $\pm$0.8{~\AA}, $\pm$1.2{~\AA}), 
\cite{shen14a} obtained the accurate Doppler velocity ranging from 6 km s$^{-1}$ to 14 km s$^{-1}$ of a winking filament without horizontal motions.
The above illustrates that the accuracy of Doppler velocities are determined by the data spectral resolution.
Benefiting form the high-resolution spectral data provided by CHASE/HIS, 
the 3D velocity of the filament motion in this study have been derived from the unambiguous observation of the simultaneous horizontal and vertical oscillations, 
which almost completely eliminated the projection effect.

It should be noted that the maximum velocity of the Doppler redshift (18 km s$^{-1}$) was slightly smaller than that of the blueshift (24 km s$^{-1}$) during the first cycle of the filament oscillation.
Excluding the calculation errors and the effect of the solar rotation,
we suppose that since the initial displacement of the filament was greater than its height, 
the downward motion of the filament should be impeded by the chromospheric material and magnetic field below (see Figure~\ref{fig5}), 
after being squeezed, the chromospheric material and magnetic field could provide gas and magnetic pressure gradient for the upward motion of the filament, 
while the resistance of the corona above can be negligible.

As we known, the direction of the filament oscillation is closely associated with the wave-filament interaction \citep{Liu2013,shen14b}, 
while the direction of the coronal-Moreton wave is determined by the inclination of the corresponding eruption or CME.
The inclination angles proposed by \cite{zheng2023} were estimated by the lowest part of coronal wave fronts with respect to the solar radial direction. 
Correspondingly, the inclination angle in this work can be estimated by the lowest part of cone model with respect to the solar radial direction.
\cite{zheng2023} concluded that the bright segments of the EUV wavefronts and the simultaneous Moreton wave should be excited by the non-radial eruptions with inclination angles of 63$^\circ-$76$^\circ$.
In this work, we found that the bright segment of the EUV wave was contemporal with the dark front of the Moreton wave.
With the addition of angular width ($\omega/2$), the inclination angle of the loops eruption in this event is determined to be 64$^\circ$ northward,
which reaches the value of the highly inclination.
Thus, the inclination angle of the loops eruption here could be sufficient to excite the coronal-Moreton wave, which in turn triggered the simultaneous horizontal and vertical filament oscillation.

\section{acknowledgements}
We thank the referee for reading the manuscript, and we greatly appreciate the valuable comments that significantly improved this paper.
We thank the CHASE, SDO and STEREO teams for providing data.
The observation data is from CHASE mission supported by China National Space Administration.
SDO is a mission of NASA\rq{}s Living With a Star Program.
AIA and HMI data are courtesy of the NASA/SDO science teams.
The authors are supported by the Special Research Assistant Project of Chinese Academy of Sciences, the Project funded by China Postdoctoral Science Foundation (2023M733734),  the Prominent Postdoctoral Project of Jiangsu Province, the Youth development program of Purple Mountain Observatory (2022000123), and Yunnan Key Laboratory of Solar Physics and Space Science under number YNSPCC202206, the National Key R\&D Program of China 2021YFA1600500 (2021YFA1600502), and the Chinese foundations NSFC 12173092, 12250014, 12333009, 12373065.



\end{document}